\documentclass[paper=A4,11pt,DIV=12,headings=big]{scrartcl}
\pdfoutput=1

\usepackage[a4paper, top=1.2in, bottom=1.2in, left=1.1in, right=1.1in]{geometry}
\usepackage{amsfonts}
\usepackage{amsmath,amssymb}
\usepackage{mathrsfs}
\usepackage[normal]{caption}

\usepackage{cancel}
\usepackage{xcolor}
\usepackage{xspace}
\usepackage{cite}
\usepackage{enumerate}
\usepackage{graphicx}
\graphicspath{{./Figs/}}
\usepackage{wrapfig}

\renewenvironment{abstract}{%
\begin{minipage}{0.95\textwidth}
}
{\par\noindent\end{minipage}}

\usepackage[multiple]{footmisc}
\let\oldfootnote\footnote\renewcommand\footnote[1]{\oldfootnote{\hspace{2mm}#1}}

\definecolor{darkblue}{rgb}{0,0,0.9}
\usepackage{hyperref}
\hypersetup{
linktocpage,
colorlinks,
citecolor=darkblue,
filecolor=darkblue,
linkcolor=darkblue,
urlcolor=darkblue
}
\allowdisplaybreaks
\addtokomafont{disposition}{\rmfamily\boldmath}

\newcommand{\mc}{\mathcal}

\newcommand{\<}{\langle}
\renewcommand{\>}{\rangle}

\newcommand{\eps}{\epsilon}

\def\bsmumu{\ensuremath{B^0_s \to \mu^{+} \mu^{-}}\xspace}
\def\mmumu{\ensuremath{m_{\mu^{+} \mu^{-}}}\xspace}
\def\bsmumugamma{\ensuremath{B^0_s \to \mu^{+} \mu^{-} \gamma}\xspace}

\def\sla#1{\setbox0=\hbox{$#1$}\dimen0=\wd0
      \setbox1=\hbox{/} \dimen1=\wd1 \ifdim\dimen0>\dimen1
      \rlap{\hbox to \dimen0{\hfil/\hfil}} #1                        \else
      \rlap{\hbox to \dimen1{\hfil$#1$\hfil}}
      /   \fi}

\newcommand{\be}{\begin{equation}}
\newcommand{\ee}{\end{equation}}
\newcommand{\bea}{\begin{eqnarray}}
\newcommand{\eea}{\end{eqnarray}}

\newcommand{\nn}{\nonumber}

\DeclareOldFontCommand{\rm}{\normalfont\rmfamily}{\mathrm}
\DeclareOldFontCommand{\sf}{\normalfont\sffamily}{\mathsf}
\DeclareOldFontCommand{\tt}{\normalfont\ttfamily}{\mathtt}
\DeclareOldFontCommand{\bf}{\normalfont\bfseries}{\mathbf}
\DeclareOldFontCommand{\it}{\normalfont\itshape}{\mathit}
\DeclareOldFontCommand{\sl}{\normalfont\slshape}{\@nomath\sl}
\DeclareOldFontCommand{\sc}{\normalfont\scshape}{\@nomath\sc}

\begin{document}

%%%% TITLE PAGE
\begin{flushright}
\small
LAPTH-044/16
\end{flushright}
\vskip0.5cm

\begin{center}
%%%% TITLE
{\sffamily \bfseries \LARGE \boldmath
\bsmumugamma from \bsmumu}\\[0.8 cm]
%%%% AUTHORS
{\normalsize \sffamily \bfseries Francesco Dettori$^a$, Diego Guadagnoli$^b$ and M\'eril Reboud$^{b,c}$} \\[0.5 cm]
\small
$^a${\em European Organization for Nuclear Research (CERN), Geneva, Switzerland}\\[0.1cm]
$^b${\em Laboratoire d'Annecy-le-Vieux de Physique Th\'eorique UMR5108\,, Universit\'e de Savoie Mont-Blanc et CNRS, B.P.~110, F-74941, Annecy-le-Vieux Cedex, France}\\[0.1cm]
$^c${\em \'Ecole Normale Sup\'erieure de Lyon, F-69364, Lyon Cedex 07, France}
\end{center}

\medskip

\begin{abstract}\noindent
The \bsmumugamma decay offers sensitivity to a wider set of effective operators than its non-radiative counterpart \bsmumu, and a set that is interesting in the light of present-day discrepancies in flavour data. On the other hand, the direct measurement of the \bsmumugamma decay poses challenges with respect to the \bsmumu one. We present a novel strategy to search for \bsmumugamma decays in the very event sample selected for \bsmumu searches. The method consists in extracting the \bsmumugamma spectrum as a ``contamination'' to the \bsmumu one, as the signal window for the latter is extended downward with respect to the peak region. We provide arguments for the actual practicability of the method already on Run-2 data of the LHC.
\end{abstract}

\vspace{1.0cm}

\renewcommand{\thefootnote}{\arabic{footnote}}
\setcounter{footnote}{0}
%%%% END TITLE PAGE

\noindent The \bsmumu decay is one of the cleanest low-energy probes of physics beyond the Standard Model (SM). The presence of new dynamics at scales above the electroweak one can be probed most generally by adopting an effective-field theory approach, whereby beyond-SM physics manifests itself as shifts to the Wilson coefficients of the operators of the $b \to s \ell \ell$ effective hamiltonian. In this context, the \bsmumu decay is sensitive to the scalar and pseudoscalar operators $\mc O_{S,P}^{(\prime)}$, and to the operator $\mc O_{10}$, defined as (see for example \cite{Bobeth:2001sq})
\bea
\label{eq:OSP10}
\mc O_{S} ~=~ \frac{\alpha_{\rm em}}{4\pi} m_b \, \bar{s} P_R b \, \bar \ell \ell~,~~~
\mc O_{P} ~=~ \frac{\alpha_{\rm em}}{4\pi} m_b \, \bar{s} P_R b \, \bar \ell \gamma_5 \ell~,~~~
\mc O_{10} ~=~ \frac{\alpha_{\rm em}}{4\pi} \, \bar{s} \gamma^\mu P_L b \, \bar \ell \gamma_\mu \gamma_5 \ell~,
\eea
with $\mc O_{S,P}^\prime$ defined from the unprimed counterparts via the replacements $P_R \to P_L$ and $m_b \to m_s$ in Eq. (\ref{eq:OSP10}). Within the SM, to an excellent approximation only the operator $\mc O_{10}$ contributes. The good agreement between the SM prediction \cite{Bobeth:2013uxa}
\be
\mc B(\bsmumu)_{\rm SM} = (3.65 \pm 0.23) \times 10^{-9}
\ee
and the current best measurement~\cite{CMS:2014xfa}
\be
\label{eq:Bsmumu}
\mc B(\bsmumu)_{\rm exp} = (2.8^{+0.7}_{-0.6})\times 10^{-9} =
(0.76^{+0.20}_{-0.18})\times \mc B(\bsmumu)_{\rm SM}\quad, 
\ee
forces scalar and pseudoscalar contributions to negligible values \cite{Alonso:2014csa,Altmannshofer:2014rta}. On the other hand $\lesssim$O(15\%) new contributions to the Wilson coefficient of the operator $\mc O_{10}$ are allowed by present errors, and actually favoured -- provided they are in destructive interference with the SM contribution -- by the about 25\% too low central value in Eq. (\ref{eq:Bsmumu}).

Adding a photon to the final state, namely considering the \bsmumugamma decay, yields an observable sensitive not only to $\mc O_{10}$, but also to $\mc O_9$ and to the electromagnetic-dipole operator $\mc O_7$, as well as to their chirality-flipped counterparts \cite{Kruger:2002gf,Geng:2000fs,Dincer:2001hu,DescotesGenon:2002ja,Aliev:1996ud,Melikhov:2004mk}. (The sensitivity to $\mc O_7$ occurs for values of the final-state invariant mass squared close to zero; as our discussion will be concerned with the high invariant-mass region, this operator will not be considered any further.) Increasing the number of observables sensitive to these operators, especially $\mc O_9$ and $\mc O_{10}$, is very important in the light of present data. In fact, the LHCb experiment as well as the $B$ factories performed a number of measurements of $b \to s$ transitions, and the overall agreement with the SM is less than perfect. Discrepancies concern in particular:
\begin{itemize}

\item the ratio $R_K$ of the branching fractions for $B^+ \to K^+ \ell^+ \ell^-$, with $\ell = \mu, e$ \cite{Aaij:2014ora}
\be
R_K \equiv \frac{\mc B(B^+ \to K^+ \mu^+ \mu^-)}{\mc B(B^+ \to K^+ e^+ e^-)}~,
\ee
showing a $2.6\sigma$ deficit with respect to the SM \cite{Bordone:2016gaq,Bobeth:2007dw,Bouchard:2013mia,Hiller:2003js};

\item the absolute $B^+ \to K^+ \mu^+ \mu^-$ branching ratio \cite{Aaij:2014pli,Aaij:2012vr}, about 30\% lower than the SM \cite{Bobeth:2011gi,Bobeth:2011nj,Bobeth:2012vn}; 

\item the measurement of $\mc B(B^0_s \to \phi \mu^+ \mu^-)$ \cite{Aaij:2015esa,Aaij:2013aln}, lower than the SM prediction by more than $3\sigma$ \cite{Aaij:2015esa};

\item the angular distribution of the $B^0 \to K^{\ast 0} \ell^+ \ell^-$ decays and, most notably, the quantity known 
as $P'_5$ \cite{Descotes-Genon:2013vna}, measured by both LHCb \cite{Aaij:2013qta,Aaij:2015oid} and Belle \cite{Abdesselam:2016llu}, whose theoretical error is, however, still debated \cite{Khodjamirian:2010vf,Descotes-Genon:2013wba,Lyon:2014hpa,Jager:2014rwa,Ciuchini:2015qxb}.

\end{itemize}
Remarkably, one can find a consistent theoretical interpretation of all these discrepancies, as well as of the last equality in Eq. (\ref{eq:Bsmumu}), within an effective-theory approach \cite{Hiller:2014yaa,Ghosh:2014awa,Altmannshofer:2014rta,Descotes-Genon:2015uva,Hurth:2016fbr}. Data can be accounted for at one stroke with new contributions to $C_{9}$ only, or jointly to $C_{9}$ and $C_{10}$. Furthermore, the indications of new-physics (NP) couplings preferring muons over electrons can be accommodated by invoking an effective interaction coupled dominantly (before electroweak-symmetry breaking) to third-generation fermions \cite{Glashow:2014iga}. This possibility would even allow to relate the mentioned $b \to s$ discrepancies with others existing in $b \to c$ transitions \cite{Bhattacharya:2014wla}.

Given its sensitivity to $C_9$ and $C_{10}$ alike, the radiative decay \bsmumugamma offers an additional probe into physics beyond the SM, and in particular a probe of couplings that are interesting in the light of current data. However, the direct measurement of radiative hadron decays is harder with respect to their non-radiative counterparts for various reasons. First, the detection and reconstruction efficiency of a photon is typically smaller than the one of charged tracks. Secondly, the energy being shared with the additional photon makes the other daughter particles softer, yielding smaller trigger and reconstruction efficiencies. Furthermore, the invariant mass reconstructed in decays with photons has, at these energies, a worse resolution than in decays without. This in turn leads to a larger background under the signal peak. The above considerations hold in particular for hadron-collider experiments, due to the high occupancy of typical events, and for low-energy processes such as those of interest to flavour physics.  Despite these difficulties, rare radiative decays with branching ratios of order $10^{-6} \div 10^{-7}$ have been observed and exploited for NP searches by several experiments, see \cite{Blake:2016olu} for a recent review. However, the rates just mentioned are still very `abundant' if compared to the \bsmumu decay and its radiative counterpart. The latter poses therefore a formidable challenge for direct detection.

In this paper we propose a method to search for \bsmumugamma events in the very same event sample selected for the $\mc B(\bsmumu)$ measurement. In one sentence, the method consists in measuring $\bsmumugamma$ as ``contamination'' to $\bsmumu$, by suitably enlarging downward the signal window for the latter search. This possibility requires a number of qualifications, since the $\bsmumu$ measurement itself comes with some subtleties as far as photons are concerned -- notably the treatment of soft final-state radiation.

\begin{figure}[t]
\begin{center}
  \includegraphics[width=0.95\linewidth]{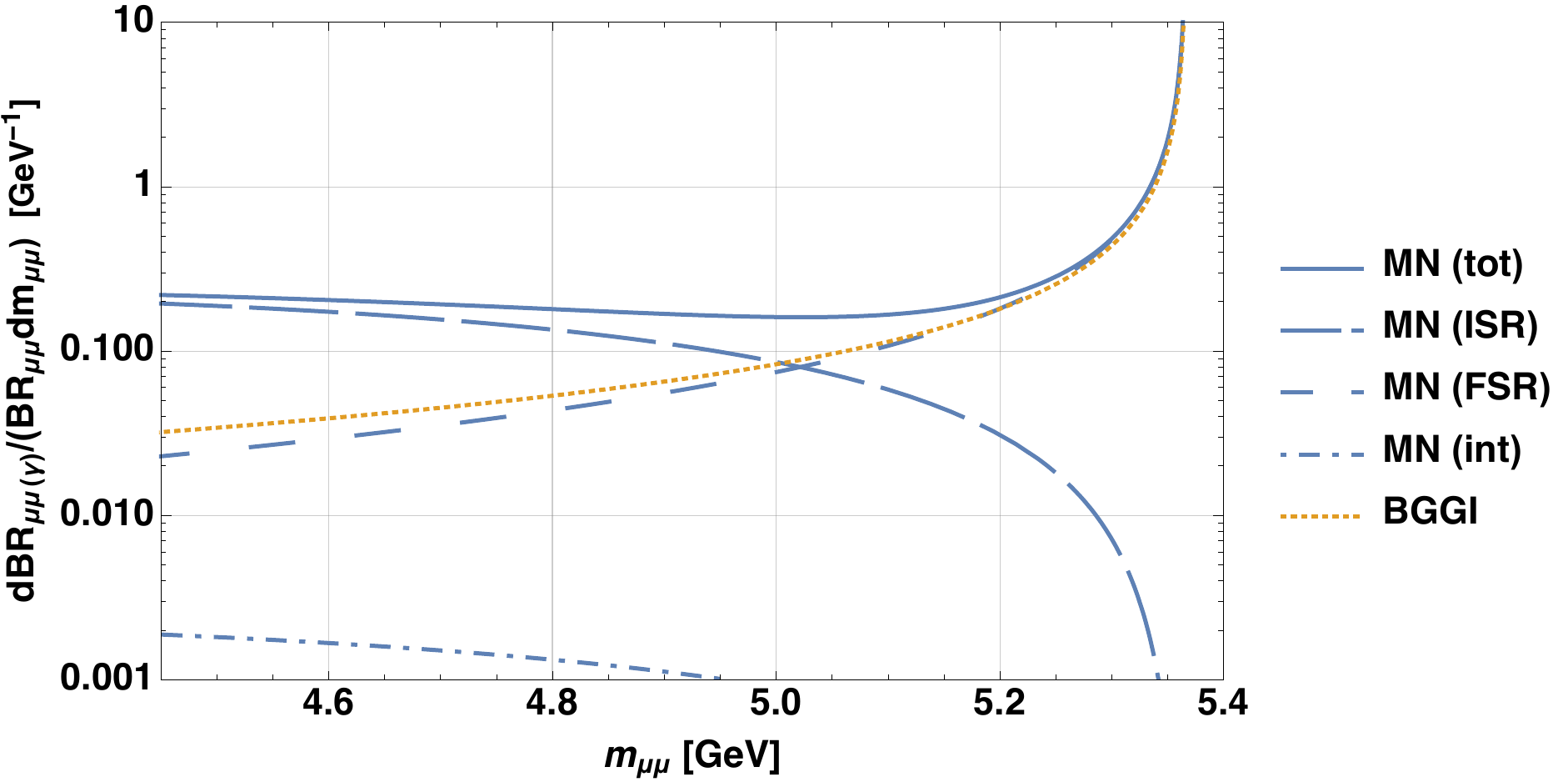} \hfill
  \caption{Breakup of the full $\bsmumugamma$ spectrum (solid blue) -- calculated in Ref. \cite{Melikhov:2004mk}, denoted as MN in the legend -- into its pure ISR component (long-dashed blue), FSR one (medium-dashed blue), and ISR-FSR interference (dot-dashed blue). We also report the $\bsmumu + n \gamma$ spectrum in the soft-photon approximation (dotted orange) from Ref. \cite{Buras:2012ru}, denoted as BGGI in the legend.}
  \label{fig:MN_ISR_FSR}
\end{center}
\end{figure}

In an idealised measurement, the $\bsmumu$ decay appears as a peak in the invariant mass squared of the two final-state muons, with negligible intrinsic width.%
\footnote{The experimental resolution in the muon momenta gives this peak an approximately Gaussian shape, the width being for example of about 25~MeV for the LHCb experiment and ranges from 32 to 75~MeV for the CMS experiment~\cite{CMS:2014xfa}.} Already at this level, however, the `definition' of the final-state muons is complicated by the fact that they emit soft bremsstrahlung, giving rise to $\bsmumu + n \gamma$ decays, with the $n$ photons undetected. This effect is however well known \cite{YFS,WeinbergIR,IsidoriIR}. As reappraised in Ref. \cite{Buras:2012ru}, it can be summed analytically to all orders in the soft-photon approximation, yielding a multiplicative correction to the non-radiative rate. This contribution skews downwards the peak region of the \bsmumu distribution, as shown by the dotted orange curve of Fig.~\ref{fig:MN_ISR_FSR}.

In order to compare the measured \bsmumu rate with the theoretical one \cite{Bobeth:2013uxa}, the mentioned soft-radiation tail due to $\bsmumu + n \gamma$ needs to be subtracted off. For example, a \bsmumu signal window extending down to about $5.3$ GeV is equivalent to a single-photon energy cut $E_\gamma \simeq 20 \div 100$ MeV, amounting to a negative shift of $\mc B(\bsmumu)$ as large as $15\%$ \cite{Buras:2012ru}.
Experimentally, the radiative tail is obtained and taken into account using Monte Carlo $\bsmumu$ events with full detector simulation and with bremsstrahlung photon emission modelled through the \textsc{Photos} application~\cite{Golonka:2005pn}. The advantage of this approach over the analytic one \cite{Buras:2012ru} is that the correction factor is already adjusted for detector efficiencies.

\begin{figure}[t]
\begin{center}
  \includegraphics[width=0.85\linewidth]{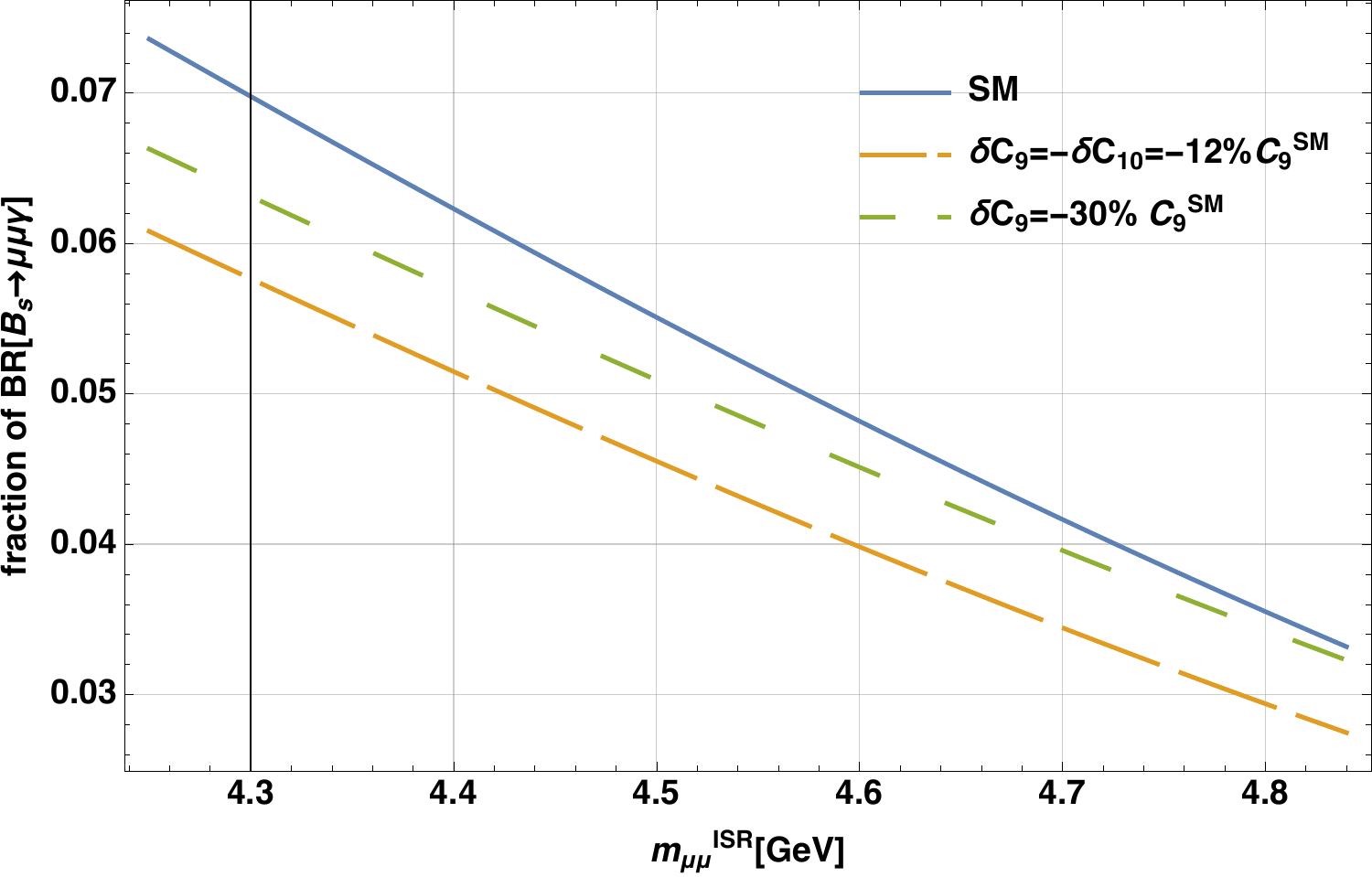} \hfill
  \caption{Fraction of the full \bsmumugamma spectrum as a function of the chosen signal-region lower bound $m_{\mu\mu}^{\rm ISR}$, for three scenarios, specified in the legend. See text for details.}
  \label{fig:BsmmgamFRAC}
\end{center}
\end{figure}
For softer and softer photons (or equivalently for $m_{\mu^+ \mu^-}$ closer and closer to the $B^0_s$ peak region), the single-photon component in $\mc B(\bsmumu + n \gamma)$ is expected to match the radiative branching ratio $\mc B(\bsmumu \gamma)$, as computed in Ref. \cite{Melikhov:2004mk} to leading order in $\alpha_{\rm em}$ (see also update in Ref. \cite{Kozachuk:2016ypz}).\footnote{In this spirit, we would also expect the ISR component of the $\bsmumugamma$ spectrum calculated in Ref. \cite{Aditya:2012im} to match, in the $m_{\mu^+ \mu^-}$ region close to the endpoint of this distribution, the corresponding spectrum calculated in Ref. \cite{Melikhov:2004mk}. We actually find that, while the two distributions have a similar shape, the distribution from \cite{Aditya:2012im} is, in the mentioned $m_{\mu^+ \mu^-}$ region, a factor of almost 4 above the one in \cite{Melikhov:2004mk}. Barring a normalisation typo in Ref. \cite{Aditya:2012im}, we are unable to physically interpret this difference.}~%
This is indeed the case, as shown by comparing the solid blue distribution with the dotted orange one in Fig.~\ref{fig:MN_ISR_FSR}. We can actually go farther in this comparison by separating the contributions due to photons emitted from final-state leptons -- to be denoted as final-state radiation (FSR) -- with respect to the rest -- to be collectively referred to as initial-state radiation (ISR) contributions. This separation makes sense to the extent that we can identify two regions in $m_{\mu\mu}$ where only one of the two contributions is dominant. The breakup of the \bsmumugamma spectrum into its different components is likewise reported in Fig. \ref{fig:MN_ISR_FSR}. As well known, the FSR contribution is dominant for soft photons (or high $m_{\mu\mu}$), whereas the ISR one dominates for harder and harder photons, namely as $m_{\mu\mu}$ decreases from the peak region. The crossover region between the two contributions is at $m_{\mu\mu} \approx 5.0$ GeV. More importantly for our purposes, the contribution from the interference term is always below 1\% of the total spectrum.\footnote{On the correct sign of the interference term see Ref. \cite{Guadagnoli:2016erb}.} This holds true fairly generally also beyond the SM. In particular, shifts in $C_9$ and $C_{10}$ with opposite sign with respect to the respective SM contributions, as hinted at by the recent $b \to s$ discrepancies mentioned earlier, tend to decrease the interference term even further. As a consequence, the ISR and FSR contributions can be treated as two basically independent spectra.

In short, to the extent that the FSR contribution can be systematically subtracted off, as is the case for \bsmumu searches, one can measure the ISR component of the \bsmumugamma spectrum -- and thereby the \bsmumugamma differential rate -- as ``contamination'' of \bsmumu candidate events as the signal window is enlarged downwards. We note that such contamination is, in principle, already present in existing \bsmumu searches. However, it is negligible in the typical window of $\pm 3 \div 5$ standard deviations around the \bsmumu peak, and its smooth distribution can be absorbed in other background distributions due, for example, to combinatorial background or partially reconstructed $B$ decays. For this reason it was typically not included as {\em separate} component in recent \bsmumu decay measurements~\cite{Aaij:2013aka,Chatrchyan:2013bka,Aaboud:2016ire}.

On the other hand, as the signal window is enlarged downwards, the ISR component of the \bsmumugamma spectrum becomes sizable. Fig. \ref{fig:BsmmgamFRAC} shows in more detail how large this contamination is expected to be. The figure displays the fraction of the full \bsmumugamma spectrum as a function of the chosen value for $m_{\mu\mu}^{\rm ISR}$ for the SM case, as well as for the two scenarios that best fit the $b \to s$ anomalies: one with a $V-A$ shift to $C_9$ and $C_{10}$, and such that $\delta C_9 = - 12\% \, C_9^{\rm SM}$, the other with a $C_9$-only shift such that $\delta C_9 = - 30\% \, C_9^{\rm SM}$ \cite{Altmannshofer:2014rta}. The figure reveals that this fraction is larger within the SM than in the considered NP scenarios. For example, it is about 4.8\% in the SM for a \bsmumu signal window extending down to $m_{\mu\mu}^{\rm ISR} = 4.6$ GeV, whereas it is about 4\% in the $V-A$ scenario.

\begin{figure}[t]
\begin{center}
\includegraphics[width = 0.95\linewidth]{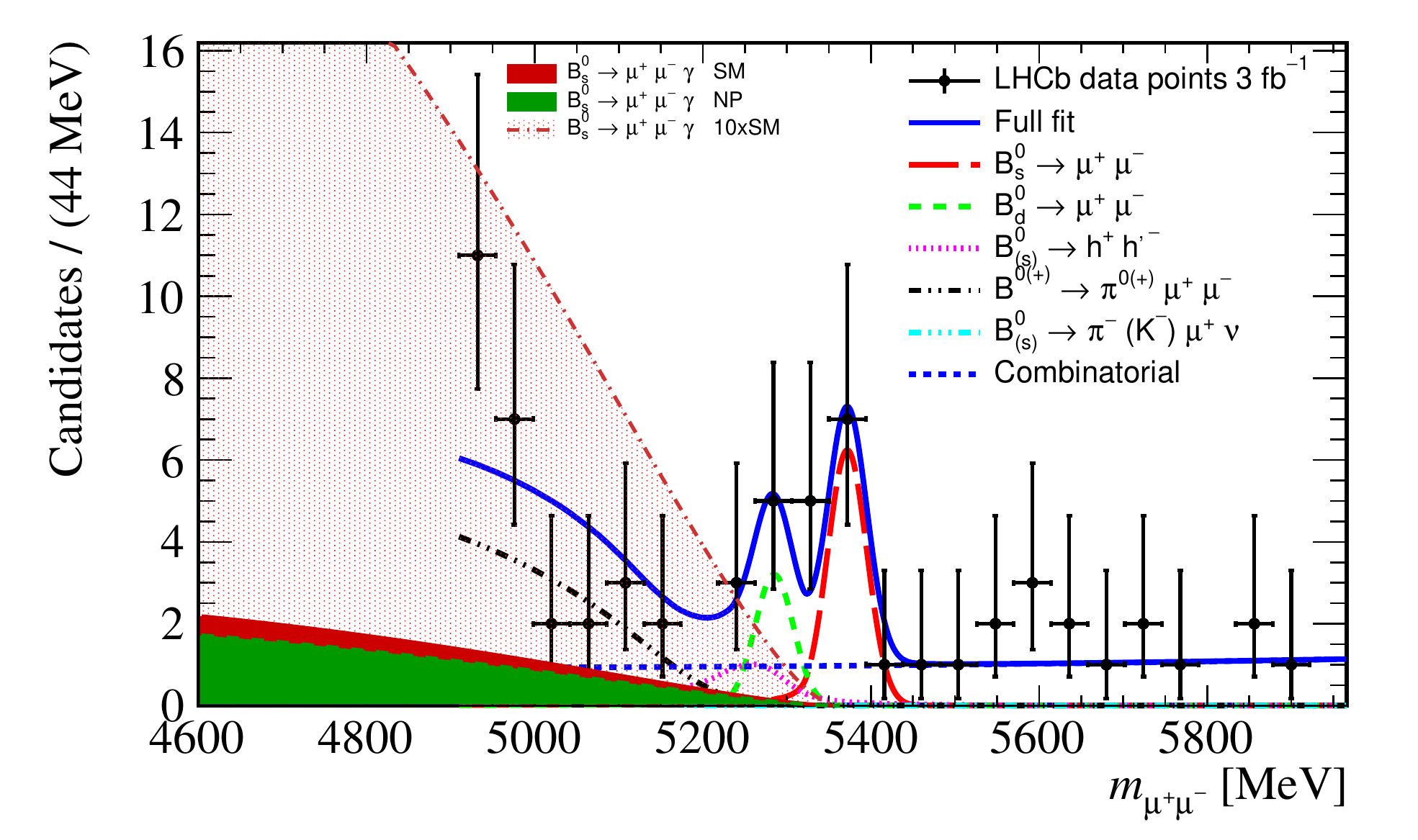}
\end{center}
\caption{Dimuon invariant mass distribution from LHCb's measurement of $\mc B(\bsmumu)$  \cite{Aaij:2013aka} overlayed with the contribution expected from \bsmumugamma decays (ISR only). Assumes flat efficiency versus \mmumu. The line denoted as `\bsmumugamma NP' refers to the $V-A$ case with $\delta C_9 = - 12\% \, C_9^{\rm SM}$ (see also Fig.~\ref{fig:BsmmgamFRAC}). The two filled curves are not stacked onto each other.
}\label{fig:bsmumulhcb}
\end{figure}
We also note that the associated event yield is large, comparable to that for the \bsmumu signal, because the \bsmumugamma rate integrates to a total branching ratio of about $2 \times 10^{-8}$ \cite{Melikhov:2004mk}, an order of magnitude above the \bsmumu one. The expected size of the \bsmumugamma spectrum is displayed in Fig.~\ref{fig:bsmumulhcb}, by superimposing this spectrum to the recent LHCb \bsmumu analysis of Ref. \cite{Aaij:2013aka}. We show the case of a SM signal as well as the NP case mentioned earlier, namely $\delta C_9 = - \delta C_{10} = - 12\% \, C_9^{\rm SM}$.
From the absolute size of these curves we can already infer that NP scenarios with the \bsmumugamma spectrum enhanced by orders of magnitude with respect to the SM are unlikely in the light of data: as shown in Fig. \ref{fig:bsmumulhcb}, a factor of 10 enhancement would result in a substantial distortion of the measured spectrum from $m_{\mu \mu} \simeq 5.1$ GeV downwards.

The \bsmumugamma spectrum shown in Fig. \ref{fig:bsmumulhcb} is obtained from our theoretical calculation, i.e. it is {\em not} a fit to existing \bsmumu data. The spectrum assumes that normalisation and efficiency be equal to those of the \bsmumu distribution itself. This is exactly true by definition at the endpoint $m_{\mu\mu} = m_{B^0_s}$, and increasingly less so for lower masses, due to the various selection criteria. For example, typical analyses enforce pointing requirements with respect to the primary interaction vertex, and the latter are less satisfied when an additional undetected photon is present. These issues can only be validated in full Monte Carlo simulations of the considered experiment and analysis.

With enough statistics, one can go beyond the integrated \bsmumugamma branching ratio, and measure the \bsmumugamma spectrum. This could be within reach of LHC experiments with Run 2 data. In fact, shifts to the differential branching ratio are roughly linear in shifts to $C_9$ or $C_{10}$. Therefore, for a $C_9$ or $C_{10}$ deviation of the order of $15\%$ (as hinted at by the global fits to $b \to s$ data), the corresponding variation in the spectrum is expected to be about $15\%$ as well. Then, a fit to data could resolve such shift at one standard deviation for an event yield of about $50$.

The above argument is of statistical nature only, i.e. it disregards systematic uncertainties. There are two prominent sources of such errors. The first is the theoretical error associated to the \bsmumugamma spectrum prediction \cite{Melikhov:2004mk}. The dominant source of uncertainty in this respect is by far the one associated to the $B^0_s \to \gamma$ vector and axial form factors, defined from the relations \cite{Melikhov:2004mk}
\bea
\label{eq:ffdef}
\< \gamma(k,\eps) | \bar s \gamma^\mu \gamma_5 b | B^0_s(p)\> &=& 
i e \, \eps^*_\nu \, (g^{\mu \nu} p k - p^\nu k^\mu) \, 
\frac{F_A(q^2)}{M_{B^0_s}}~,\nn \\
\< \gamma(k,\eps) | \bar s \gamma^\mu b | B^0_s(p)\> &=& 
e \, \eps^*_\nu \, \eps^{\mu \nu \rho \sigma} p_\rho k_\sigma \, \frac{F_V(q^2)}{M_{B^0_s}}~.
\eea
To the authors' knowledge, no first-principle calculation of these form factors exists, for example within lattice QCD. The form-factor predictions used in this work are obtained from the recent analysis \cite{Kozachuk:2015kos} of heavy-meson transition form factors, based on the relativistic constituent quark model \cite{Anisovich:1996hh,Melikhov:2001zv}. The analytic expressions for the form factors from the constituent quark model reproduce the known results from QCD for heavy-to-heavy and heavy-to-light form factors \cite{Melikhov:2000yu}. Form-factor predictions within this model are thereby attached an uncertainty of about 10\%, implying a 20\% uncertainty on the branching-ratio prediction. It is clear that such level of accuracy is not sufficient to clearly resolve the effects expected from new physics (see legend of Fig.~\ref{fig:BsmmgamFRAC}). However, what is needed for the proposed method are the form factors in the high-$q^2$ range close to the kinematic endpoint. This range is the preferred one for lattice-QCD simulations.

The second potential source of systematic uncertainty for our method is of experimental nature. The impact of this uncertainty depends on the actual possibility to well constrain the other background components populating the signal window as it is enlarged towards lower values. 
This part of the spectrum, in addition to combinatorial background, consists mainly of semileptonic decays in the form $B \to h^{\pm} \mu^{\mp} \nu (+X)$, where $h$ is a pion or kaon misidentified as muon and $X$ can be any other possible hadron (not reconstructed), and rare decays such as $B^{0,+} \to h^{0,+} \mu^+ \mu^-$, which do not need any misidentification. 
While the semileptonic decays do not represent a problem as they can be constrained from control channels directly in data, the rare decays need to be estimated with a combination of experimental measurements and theoretical inputs; as an example the $B^0 \to \pi^0 \mu^+ \mu^-$ decay is not yet observed experimentally and is currently constrained using the spectral shape measured from the $B^+ \to \pi^+ \mu^+ \mu^-$ decay and theoretical estimates of the ratio of the two branching fractions~\cite{Aaij:2013aka,CMS:2014xfa}. The specific details on how to treat the single sources of backgrounds will have to be addressed by the single experiments depending on the experimental capabilities, but we do not foresee these to be irreducible backgrounds. 

We emphasize that our proposed method is potentially applicable to several other decays --~in principle the radiative counterpart of any two-body decay whereby the initial-state meson mass is completely reconstructible. Straightforward examples are provided by all the other $B_{q} \to \ell^+ \ell^- \gamma$ modes, for which the only existing limits concern $B^0 \to e^+ e^- \gamma$ or $\mu^+ \mu^- \gamma$ with a technique based on explicit photon reconstruction \cite{Aubert:2007up}. Serious consideration of these decays will be timely when mature measurements of the corresponding non-radiative decays will become available.

In conclusion, we presented a novel method for the extraction of the $\bsmumugamma$ spectrum at high $m_{\mu\mu}^2$. The method avoids the drawbacks of explicit photon reconstruction, and takes advantage of the fact that this spectrum inevitably contaminates the $\bsmumu$ event sample as the $m_{\mu\mu}^2$ signal window is enlarged downward. Fig. \ref{fig:bsmumulhcb} shows that order-of-magnitude enhancements of the \bsmumugamma decay rate are unlikely, already in the light of existing data below $m_{\mu \mu} \simeq 5.1$ GeV. More likely, the measurement will involve a dedicated fit by experiments, and this is where our method may make the difference. This method can realistically be applicable in LHC Run 2 data, and would thereby allow to set the first limit for $\mc B(\bsmumugamma)$, or provide the first measurement thereof. 

\section*{Acknowledgements}
The work of DG is partially supported by the CNRS grant PICS07229. The authors are indebted to Dmitri Melikhov for many clarifications on Ref. \cite{Melikhov:2004mk} and related work. The authors also thank Gino Isidori for comments on the manuscript, and Mikolaj Misiak for discussions.

\providecommand{\href}[2]{#2}\begingroup\raggedright\endgroup

\end{document}